**Superconductivity of the New Medium-Entropy Alloy $V_4Ti_2W$ with a Body-Centered Cubic Structure**

*Kuan Li*[a], *Weijie Lin*[b,c], *Ruixin Guo*[b,c], *Shu Guo*[b,c], *Lingyong Zeng*[a], *Longfu Li*[a], *Peifeng Yu*[a], *Kangwang Wang*[a], *Chao Zhang*[a], *Huixia Luo*[a*]

[a] School of Materials Science and Engineering, State Key Laboratory of Optoelectronic Materials and Technologies, Guangdong Provincial Key Laboratory of Magnetoelectric Physics and Devices, Key Lab of Polymer Composite & Functional Materials, Sun Yat-Sen University, No. 135, Xingang Xi Road, Guangzhou, 510275, P. R. China

[b] Shenzhen Institute for Quantum Science and Engineering, Southern University of Science and Technology, Shenzhen 518055, China;

[c] International Quantum Academy, Shenzhen 518048, China

**ABSTRACT**

Medium- and high-entropy alloy (MEA and HEA) superconductors have attracted considerable interest since their discovery. This paper reports the superconducting properties of ternary tungsten-containing MEA $V_4Ti_2W$ for the first time. $V_4Ti_2W$ is a type II superconductor with a body-centered cubic (BCC) structure. This structure belongs to the space group $Im\overline{3}m$ (No. 229) with the cell parameters $a$= $b$ = $c$ = 3.10673(14) Å. Experimental results of resistivity, magnetization, and heat capacity indicate that the superconducting transition temperature of the MEA $V_4Ti_2W$ is roughly 5.0 K. The critical magnetic fields at the upper and lower ends are 9.93(2) T and 40.7(3) mT, respectively. Interestingly, few BCC MEA superconductors with VEC greater than 4.8 have been found. The addition of tungsten leads to a VEC of 4.83 e/a for $V_4Ti_2W$, which is rarely higher than the 4.8 value. Adding tungsten element expands the variety of MEA alloys, which may improve the microstructure and mechanical properties of materials and even superconducting properties. This material could potentially offer a new platform for the investigation of innovative MEA and HEA superconductors.

## I. Introduction

Medium-entropy alloys (MEAs) and high-entropy alloys (HEAs) are alloyed materials composed of three or more elements, where the content of each element ranges from about 5 % to 35 % [1]. According to the magnitude of the desired mixing entropy ($\Delta S_{mix}$), MEAs have an R value in the range of 0.69 to 1.60 [2,3], and HEAs have an R value of 1.60 or more [4-6]. MEAs and HEAs with unique atomic arrangement and interaction mechanisms show great potential for applications in materials science, catalyst design, energy storage, and other fields [2,7-14].

Interestingly, an attractive property of MEAs and HEAs is superconductivity. Since the body-centered cubic (BCC) HEA $Ta_{34}Nb_{33}Hf_8Zr_{14}Ti_{11}$ was discovered in 2014 [15], the number of reports of MEA and HEA superconductors has been increasing. Among them, MEAs, as a structurally diverse material system with complex compositional and structural features, offer new possibilities to realize superconductivity. In line with the VEC stipulations of Matthias' rule, BCC MEA and HEA superconductors with either 4 or 5 primarily consist of elements such as Ti, Zr, V, Nb, Mo, Ta, and Hf. These materials' superconducting transition temperature ($T_{sc}$) typically shows a broad peak around a VEC of 4.5. BCC-structured MEA and HEA superconductors with a VEC greater than 4.8 are seldom observed. A VEC greater than 4.8 can be achieved using combinations of Cr, Mo, and/or W, along with 4*d* or 5*d* transition metals. However, the number of BCC MEA superconducting materials containing the element W is very limited, which makes it a valuable research topic. [1].

According to Matthias rule, many MEA superconductors composed of transition metals can be considered intermediate materials between transition metal crystalline superconductors and amorphous superconductors. TiHfNbTa is an ultra-strongly coupled *s*-wave superconductor. The bulk $T_{sc}$ of TiHfNbTa is about 6.75 K, the upper critical field ($\mu_0 H_{c2}$) is 10.46 T, and the lower critical field ($\mu_0 H_{c1}$) is 45.8 mT [14]. Interestingly, The phase transition temperature of TiVNbTa type II superconductors in the BCC structure is about 4.65 K, but there is no strong coupling [16]. In addition, due to the synergistic effect of the Debye temperature ($\Theta_D$) near the Fermi energy level and the change in the electronic energy band structure, the TiZrHfNb alloy exhibits a significantly pressure-dependent high electrical resistance, which decreases linearly by 12.5% when the pressure is increased to 5.5 GPa [17]. The $T_{sc}$ range of membrane electrodes TiZrNb, TiZrNbHf, and HEA TiZrNbHfTa is 6.4 - 8.4 K, with an estimated $\mu_0H_{c2}$ values were higher than 10 T. Hydrogenation of alloys suppressed superconductivity due to a high concentration of disordered hydrides (defects) produced in the preparation method. The superconducting states were not observed above 2 K. The superconducting state is not observed above 2 K [18]. Interestingly, the first HEA $[TaNb]_{0.31}(TiUHf)_{0.69}$ to contain electrons from actinide ions, exhibits phonon-mediated superconductivity, with a transition temperature of $T_{sc} \approx 3.2$ K and an upper critical field of $\mu_0H_{c2} \approx 6.4$ T. This extends this class of materials to actinides, and opens the way to the synthesis of superconducting materials with various radioisotopes [19]. It is worth noting that commercial niobium titanium alloys exhibit exceptional zero resistance superconductivity at an ambient pressure of 261.7 GPa despite a 45% reduction in volume, with $T_{sc}$ increasing from ≈ 9.6 to ≈ 19.1 K and no structural phase transition [20-22]. On the other hand, the BCC MEA superconductors with a valence electron count (VEC) greater than 4.8 are rarely reported.

Inspired by all these discoveries, we actively explored a new MEA superconductor $V_4Ti_2W$ with VEC = 4.83 and $\Delta S_{mix}$ = 0.9555 R. The crystallographic structure and superconducting properties were investigated in detail. The MEA $V_4Ti_2W$ is a type II superconductor in bulk form, with a $T_{sc}$ approximating 5.57 K, as evidenced by resistivity, magnetization, and specific heat measurements.

## II. Experimental Details

$V_4Ti_2W$, which is composed of vanadium (99%, 325 mesh, Aladdin), titanium (99.8%, 300 mesh, Macklin), and tungsten (99.95%, 325 mesh, Macklin), was fabricated using the arc-melting technique. The 300 mg mixture of elemental powder was pressed into pellets at a molar ratio of V: Ti: W = 4:2:1. This pellet with tungsten powder on the top and a combination of vanadium and titanium powders at the bottom was subsequently arc melted in a 0.45 atm argon environment. The ingot was repeatedly flipped and remelted to achieve homogeneity.

The confirmation of the crystal structure was conducted via Powder X-ray diffraction (PXRD). The data of PXRD was collected between 10 ° to 100 ° with a step size of 0.01 ° using a Rigaku MiniFlex (Cu K$\alpha$ radiation) with a constant scanning speed of 1 °/min at room temperature. XRD patterns underwent refinement using the Rietveld method, utilizing the FULLPROF suite software package. Additionally, the samples' morphology and microstructure were characterized using a scanning electron microscope (SEM) and energy-dispersive X-ray spectroscopy (EDS) (EVO, Zeiss). Furthermore, a physical property measurement system (PPMS, DynaCool, Quantum Design, Inc.) was utilized to measure specific heat, zero-field-cooled (ZFC) magnetization, and resistivity. The resistivity of rectangular samples (4.5 × 1.3 × 0.6 $mm^3$) was assessed through the utilization of the four-point probe technique employing PPMS. The specific heat within the temperature range of 1.9 - 15 K can be evaluated using the PPMS.

## RESULTS AND ANALYSIS

The bulk PXRD reflexes and correlation analysis of the MEA $V_4Ti_2W$ sample are shown in Fig. (1a). The PXRD reflexes indicate that the product is a pure phase. The slight expansion of the peak signals can be attributed to the significant level of disarray. Each detected PXRD reflex was identified by its corresponding Miller indices. Moreover, the data of PXRD were readily indexed to the BCC structure (space group $Im\bar{3}m$, No. 229). The lattice parameters of $V_4Ti_2W$ obtained from the PXRD profile fitted by Rietveld refinement are $a = b = c$ = 3.10673(14) Å. The refinement parameters $R_{wp}$, $R_p$, $R_e$, and $\chi^2$ are 4.91 %, 3.19 %, 3.58 %, and 1.8796, respectively. Fig (1b) shows the crystal structure of the MEA $V_4Ti_2W$ superconductor. Three elemental atoms V, Ti, and W, randomly occupy a lattice site. Furthermore, EDS was employed to analyze the uniformity and chemical makeup of the $V_4Ti_2W$ compound. The ratios of the constituent elements are shown in Fig. (1c), and the proper chemical formula is $V_{3.9}Ti_{2.2}W_{0.9}$, with the actual elemental ratios close to the design values. Meanwhile, the EDS of Fig. (1d) shows that the elemental distribution of the polycrystalline samples is homogeneous.

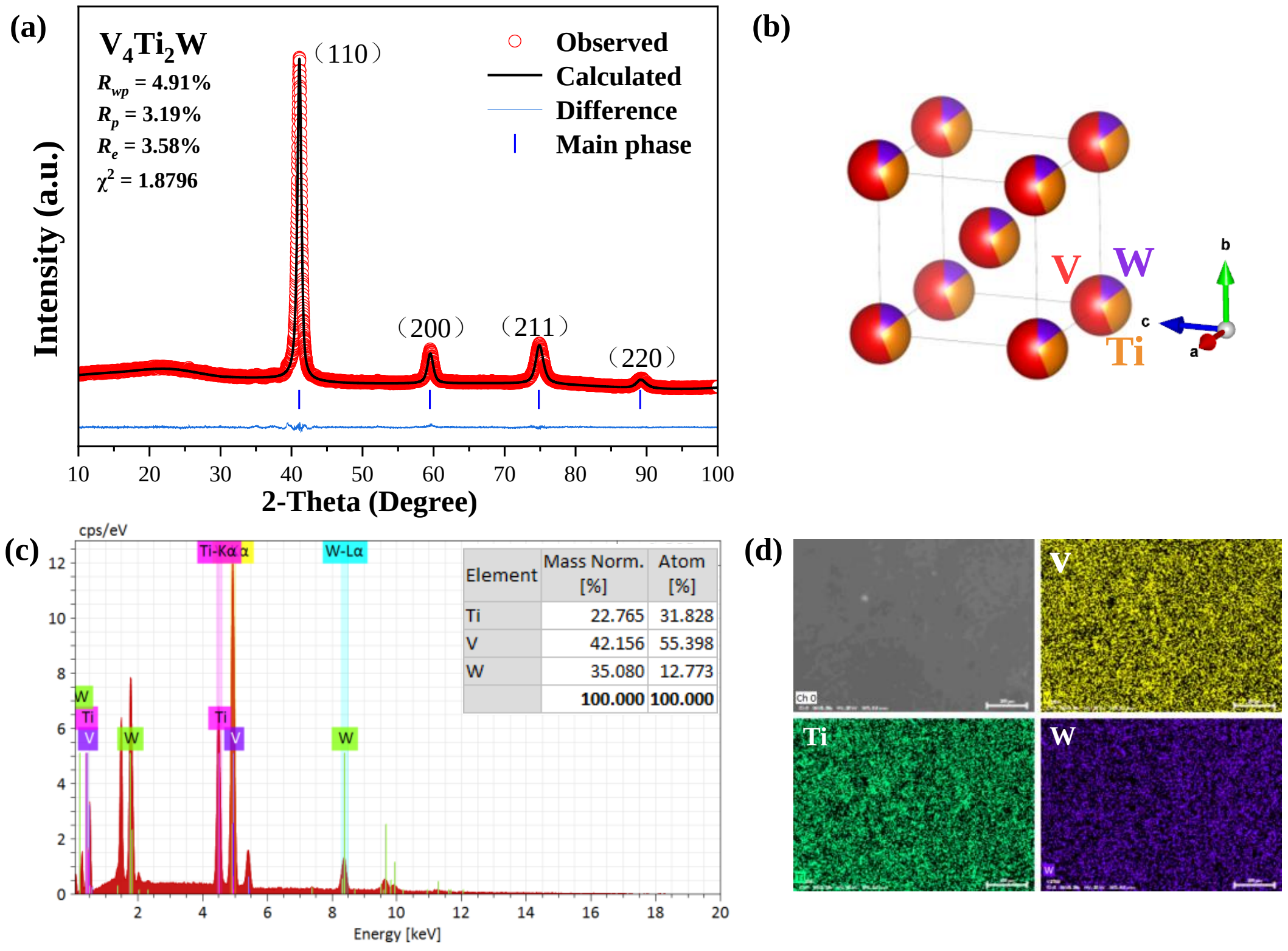

**Fig. 1.** (a) PXRD refinement of $V_4Ti_2W$. (b) Crystal structure of $V_4Ti_2W$ alloy with $Im\bar{3}m$ space groups. (c) EDS spectrum of the $V_4Ti_2W$ MEA sample, the inset shows the elemental ratio of $V_4Ti_2W$. (d) SEM image and EDS elemental mappings of the $V_4Ti_2W$ MEA sample.

The resistivity test was used to measure superconductivity. Fig. (2a) shows the zero-field resistivity measurement of $V_4Ti_2W$ MEA samples in the range of 300 ~ 1.8 K. $V_4Ti_2W$ MEA's resistivity decreases with temperature from 300 K to about 6 K, exhibiting metallic behavior. In addition, the superconductivity exhibits a sharp decrease in $\rho$(T) down to $\rho$ = 0 at low temperatures. Fig. (2b) shows the resistivity trend at temperatures of 8 ~ 1.8 K. $T_{sc}$ starts at about 5.75 K, while it reaches zero resistivity with superconducting properties at about 5.24 K. The resistivity transition of the superconducting material is very sharp at low temperatures. The abrupt change in resistivity occurs within a narrow temperature span of 0.51 K. The transition temperature of the $\rho$(T) data is $T_{sc}$ = 5.57 K, where $T_{sc}$ represents a 50 % reduction in resistivity relative to the value in the normal state. Furthermore, the residual resistivity ratio ($RRR = R_{300K}/R_{8K}$) is defined at 300 K and 8 K. For the $V_4Ti_2W$ MEA sample, the *RRR* value is 1.35.

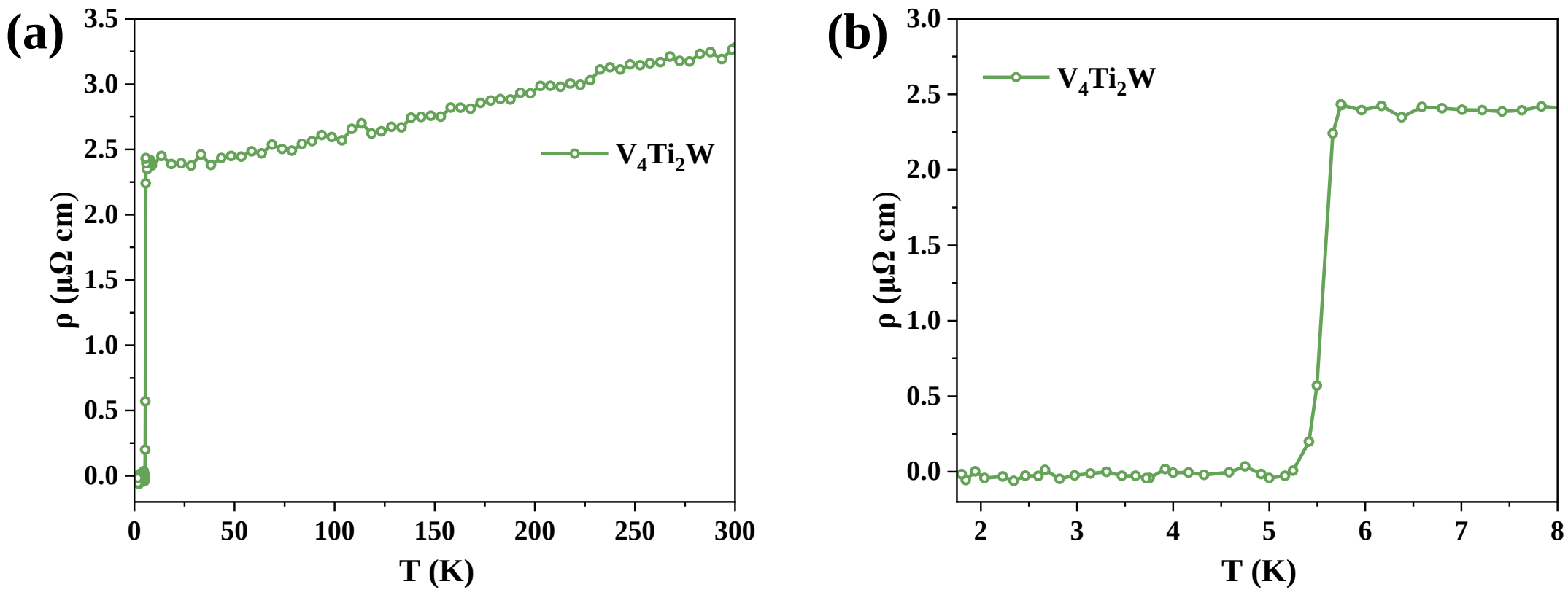

**Fig. 2.** (a) Temperature dependence of zero-field resistivity measurement of the $V_4Ti_2W$ sample. (b) Low-temperature behavior of $V_4Ti_2W$ sample near $T_{sc}$.

Fig. (3a) presents the magnetization data procured during the zero-field cooling (ZFC) and field cooling (FC) procedures under 3 mT and 10 mT magnetic fields. The magnetization data presented in this study is defined as $\chi_v = \frac{M}{H}$, where *M* denotes the volume magnetic susceptibility, and H denotes the applied magnetic field. Detecting a pronounced diamagnetic signal beneath the critical temperature $T_{sc}$ = 5.28 K validates the existence of superconductivity in the $V_4Ti_2W$ MEA specimen. The value of $4\pi\chi_v$ (1-N) (ZFC) approaches -1 at 1.8 K, indicating a Meissner volume fraction close to 100%. The isothermal magnetization M(H) data were further corrected by measuring at 1.8 K and using the demagnetization factor N to obtain a more accurate magnetization curve. The magnetization curve was then adjusted for the magnetization of the $V_4Ti_2W$

MEA sample. Within the temperature range of 1.8 to 4.0 K, Fig. (3b) displays the measured magnetic isotherms. Assuming perfect field response, the magnetization data at 1.8 K in the low-field region can be represented by a linear fit equation: $M_{fit}$ = a + bH. [23] The formula -n = $\frac{1}{4\pi(1-N)}$ is subsequently utilized to compute the demagnetization factor N. The value of N is determined to be 0.18 based on the slope (n). Fig. (3c) illustrates the M-$M_{fit}$ curves at varying temperatures. The black dashed line depicted in Fig. (3c) corresponds to the value of 1.45 emu/cm$^3$, which is derived from the data points shown in Fig. (3d). It should be noted that the value of 1.45 emu/cm$^3$ at 1.8 K is less than 2 % of M, allowing for the precise calculation of the critical field $\mu_0 H_{c1}(0)$. Fitting the extracted points to the empirical formula $\mu_0 H^*_{c1}(\mathrm{T}) = \mu_0 H^*_{c1}(0)(1\text{-}(T/T_c)^2)$ yields $\mu_0 H_{c1}(0)$ as 40.7(3) mT. Considering the demagnetization factor N, $\mu_0 H_{c1}(0) = \mu_0 H_{c1}{}^*(0)/(1\text{-N})$ yields $\mu_0 H_{c1}(0)$ = 49.6(3) mT for the $V_4Ti_2W$ MEA sample.

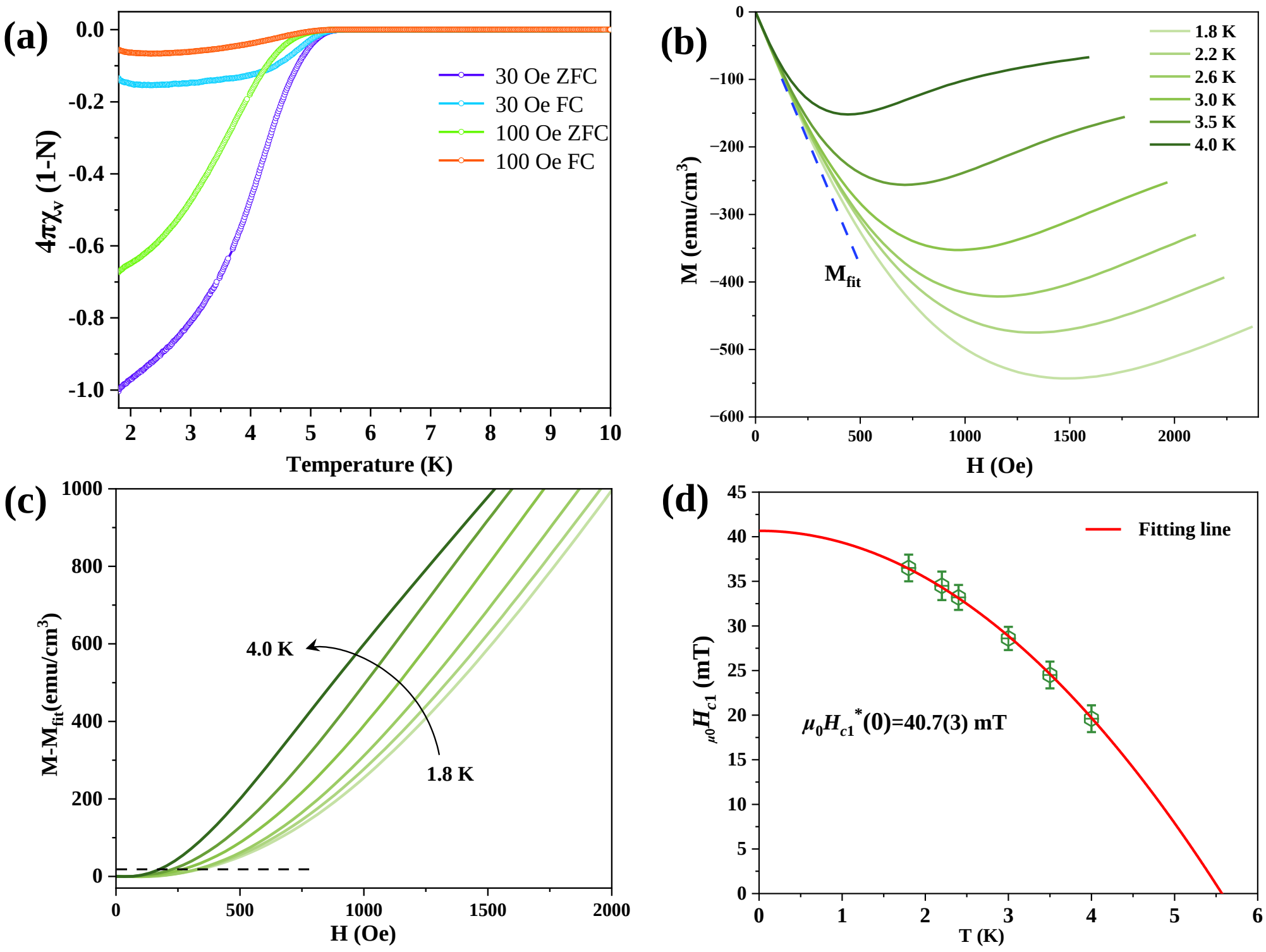

**Fig. 3.** (a) The temperature-dependent ZFC and FC magnetization of the $V_4Ti_2W$ MEA sample in various external fields. (b) The isothermal magnetization curve of the $V_4Ti_2W$ MEA within the range of 1.8 to 4 K. (c) The M -$M_{fit}$ curve measured as a function of the magnetic field in the range of 1.8 to 4.0 K. (d) The temperature dependence of the effective lower critical field of the $V_4Ti_2W$ MEA sample.

Further, the resistivity data of the $V_4Ti_2W$ superconductor under various fixed magnetic fields was investigated using the R-T measurement system, as shown in Fig. (4a). We also used the Werthamer-Helfand-Hohenberg (WHH) formula: $\mu_0H_{c2}(0) = -0.693T_{sc}(\frac{d\mu_0H_{c2}}{dT})|_{T = Tsc}$ to determine the upper critical field $\mu_0H_{c2}(0)$. The dirty limit $\mu_0H_{c2}(0)$ was computed to be 9.93(2) T employing $T_{sc}$ = 5.57 K (50 % $\rho_N$) as shown in Fig. (4b). It should be noted that the upper critical field computed by the WHH model has to be less than the Pauli limiting field $\mu_0H^{\mathrm{Pauli}}$ = 1.86 $T_{sc}$, which originates from the Pauli limiting effect. In this case, the Pauli limiting transfer field, $\mu_0H^{\mathrm{Pauli}}$ = 1.86 $T_{sc}$ = 18.47(1) T. [24] The estimated $T_{sc}$ values under different magnetic fields can be fitted to a straight line, giving a slope of $d\mu_0H_{c2}/dT$ = -1.7760. We calculated $\mu_0H_{c2}(0)$ using the Ginzburg-Landau (GL) equation: $\mu_0H_{c2}(T) = \mu_0H_{c2}(0) \times \frac{1-(T/T_c)^2}{1+(T/T_c)^2}$. Fitting the data gives $\mu_0H_{c2}(0)^{GL}$ = 8.54(1) T. The $\mu_0H_{c2}$-$T$ phase diagram of $V_4Ti_2W$ is shown in Fig. (4c).

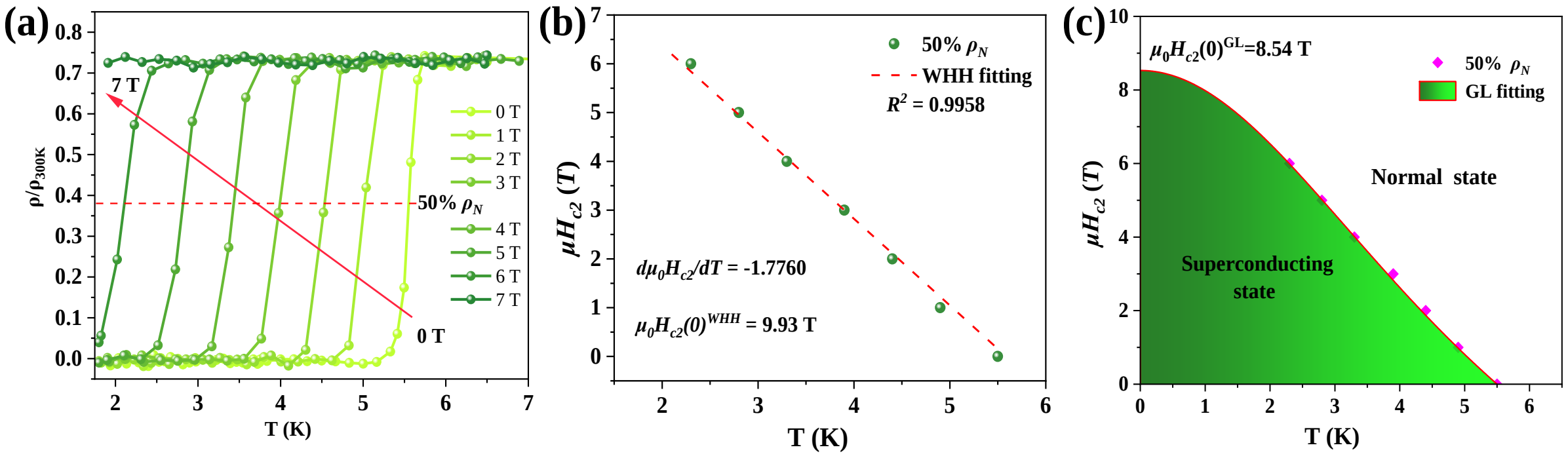

**Fig. 4**. (a) The resistivity transition under different magnetic fields from 0 to 7 T. (b) The temperature dependence of the $\mu_0 H_{c2}$ and the WHH model fitting. (c) The $\mu_0 H_{c2}$ - T phase diagram of the MEA $V_4Ti_2W$ superconductor.

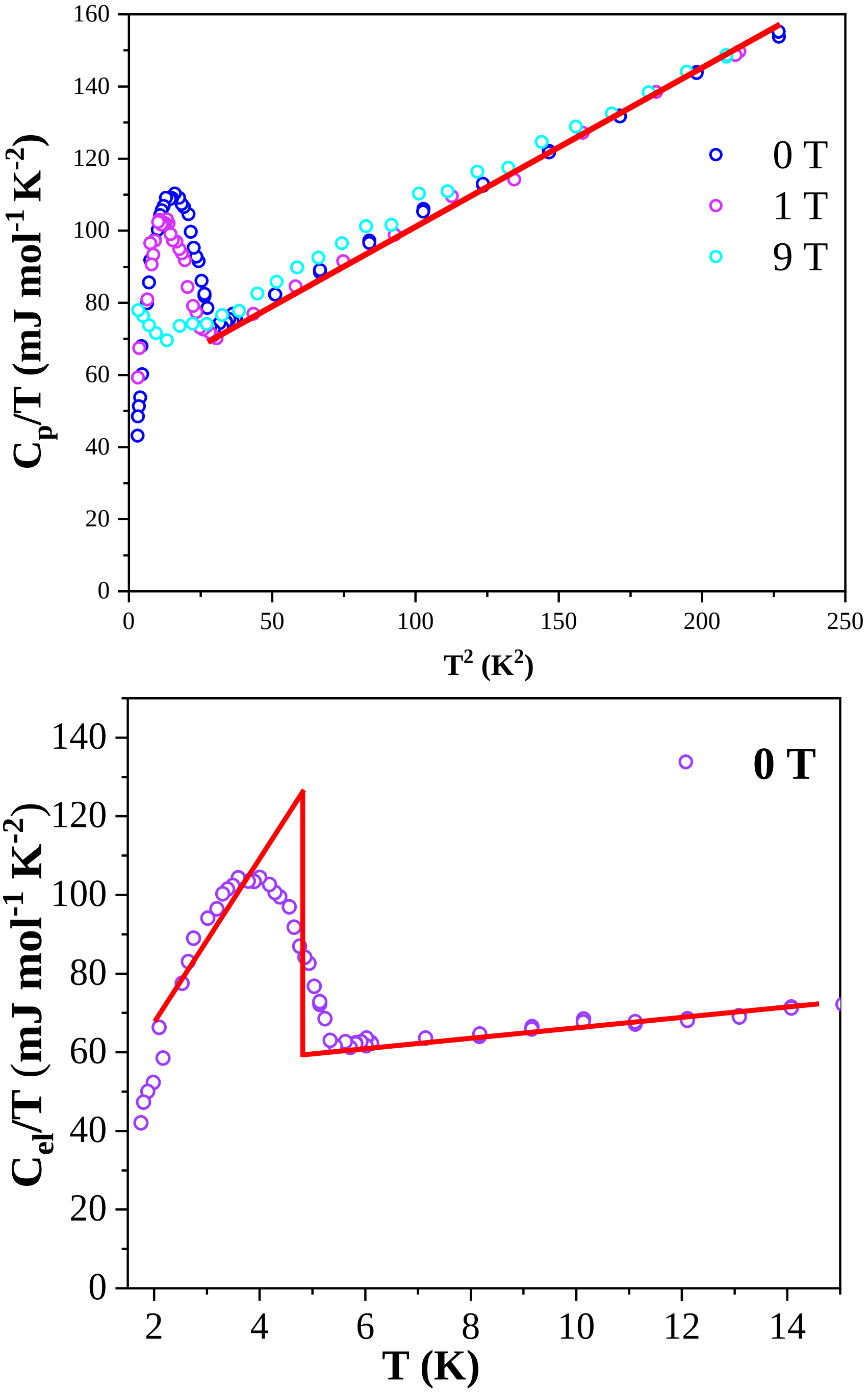

**Fig. 5**. Specific heat curves of $V_4Ti_2W$ MEA. (a) Specific heat divided by temperature $C_p$/T vs squared temperature $T^2$ under diverse applied magnetic fields. (b) Electron contribution to heat capacity as a function of $C_{el}$/T at zero magnetic field.

Heat capacity evaluations at a 0 T,1 T and 9T magnetic field corroborate the bulk superconducting condition in $V_4Ti_2W$ MEA, as depicted in Fig. (5a). In the normal state above $T_{sc}$, the experimental point can be fitted by the equation $C_p/T = \gamma_n + \beta T^2$, where $\beta$ denotes the specific heat coefficient of the lattice part, and $\gamma_n$ denotes the Sommerfeld constant of the normal state. The red line shows the estimated values of $\gamma_n$ = 60.75 mJ mol$^{-1}$ K$^{-2}$ and $\beta$ = 0.418 mJ mol$^{-1}$ K$^{-4}$ obtained from this equation, as shown in Fig. (5a). Using the Debye model and the formula $\Theta_D = (12\pi^4 nR/5\beta)^{1/3}$ to estimate the value of the Debye temperature $\Theta_D$ as 171 K, where $n$ represents the number of atoms in the unit cell, and $R$ represents the gas constant. Beyond that, Fig. (5b) shows the $C_{el}/T_{sc}$ vs $T_{sc}$ curve in the temperature range between 1.8 and 15 K under zero magnetic fields. The estimated $T_{sc}$ = 5.0 K determined by an equal-area entropy construction is consistent with the $T_{sc}$'s extracted from resistivity and magnetic susceptibility measurements for the $V_4Ti_2W$. Taking into account the conservation of entropy at $T_{sc}$, the transition temperature in the bulk phase is ascertained to be 5.0 K at zero field, aligning with the $T_{sc}$ value derived from magnetization and resistivity data. In the conserved structure shown in Fig. (4b), the normalized specific heat jump at $T_{sc}$, $\Delta C_{el}/\gamma_n T_{sc}$ ~ 1.07, is significantly smaller than the BCS value of 1.43. Therefore, $V_4Ti_2W$ MEA can be categorized as a Cooper-paired weakly coupled superconductor. Additionally, utilizing the semi-empirical McMillan formula $\lambda_{\mathrm{ep}}=\frac{1.04+\mu^* \ln\left(\frac{\Theta_{\mathrm{D}}}{1.45T_c}\right)}{\left(1\text{-}1.62\mu^*\right)\ln\left(\frac{\Theta_{\mathrm{D}}}{1.45T_c}\right)\text{-}1.04}$, [25] an estimation for the electron-phonon coupling constant $\lambda_{ep}$ of $V_4Ti_2W$ MEA has been obtained. The Coulomb pseudopotential parameter $\mu^*$ = 0.13, [14,26,27] is effective for intermetallic superconductors. Using $T_{sc}$ and $\Theta_D$ as references, we determined that the $V_4Ti_2W$ MEA has a $\lambda_e$ value of 0.81, placing it in the category of superconductors with moderate electron-phonon coupling.

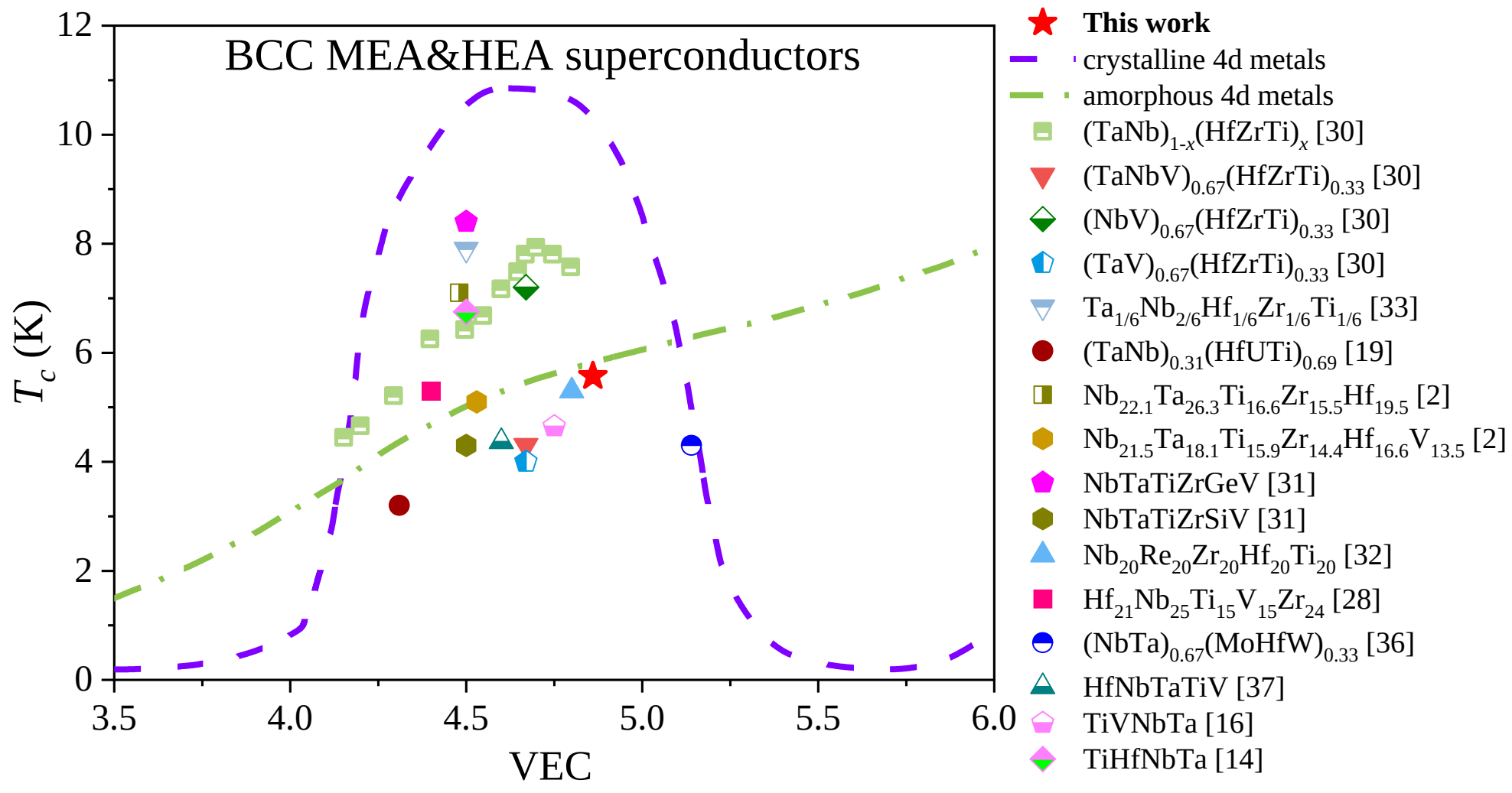

**Fig. 6**. Critical temperature $T_{sc}$ with VEC of the BCC-type MEA and HEA superconductors, crystalline 4*d* metals (purple dotted line), and amorphous 4*d* metals (green dotted line).

Fig. 6 shows a plot of $T_{sc}$ and VEC for BCC-type MEA and HEA superconductors. [2,14,16,27-37] For comparison, trend lines are also depicted for the observed $T_{sc}$ of amorphous 4*d* metal (green dashed line) and crystalline 4*d* metal (purple dashed line). The $T_{sc}$ and VEC of $V_4Ti_2W$ are comparable to other reported BCC-type MEA superconductors. As depicted in Fig. 6, the temperature range for $T_{sc}$ is observed to be distributed between 2.8 and 7.8 K for VEC = 4.67. Considering the substantial impact of the elemental composition on $T_{sc}$, Table S1 (Supporting Information) consolidates the MEA and HEA parameters for single-phase BCC structures with a $T_{sc}$ surpassing 5 K as demonstrated in Fig 6. VEC and elemental composition affect the BCC structures' $T_{sc}$ of MEA and HEA. In this paper, the VEC of $V_4Ti_2W$ is 4.86, and $T_{sc}$ = 5.57 K. The BCC MEA and HEA superconductors have highly disordered atomic arrangements, and their superconducting properties are significantly affected by the elemental compositions and the VEC. MEA and HEA superconductors possess unique mechanical and chemical properties that allow them to be used in harsh environments. Hence, we believe there is boundless potential for investigating new physical phenomena in MEA and HEA superconductors.

## CONCLUSIONS

In conclusion, we have analyzed the crystal structure and superconducting properties of the novel superconductor $V_4Ti_2W$. The compound of $V_4Ti_2W$ adopts a BCC-type structure (space group $Im\bar{3}m$, No. 229). Adding tungsten with a 5*d* orbit results in a VEC of 4.83 e/a. Adding tungsten can expand the types of MEAs and HEAs, thereby improving the microstructure and mechanical properties of the material and even the superconductivity. Our study shows that $V_4Ti_2W$ is a type-II superconductor with $T_{sc}$ ~ 5.57 K, $\mu_0H_{c2}(0)$ ~ 9.93(2) T and $\mu_0H_{c1}(0)$ ~ 40.7(3) mT. This material provides a new platform for studying new MEA and HEA superconductors. To our knowledge, the number of ternary MEAs containing tungsten elements is tiny, which finds a direction to search for new MEA and HEA superconducting materials. Low-temperature specific heat measurements show that the alloy is a conventional *s*-wave superconductor with a moderately coupled superconductor. Additionally, $V_4Ti_2W$ presents a promising opportunity for investigating the distinctive crystal structure and superconducting behavior, thereby offering valuable insights into the physical characteristics of medium and high-entropy superconductors.

**Notes**

The authors declare no competing financial interest.

## ACKNOWLEDGMENT

This work is supported by the National Natural Science Foundation of China (12274471, 11922415, 22205091), Guangdong Basic and Applied Basic Research Foundation (2022A1515011168). The experiments reported were conducted at the Guangdong Provincial Key Laboratory of Magnetoelectric Physics and Devices, No. 2022B1212010008.

## REFERENCES

[1] J. Kitagawa, S. Hamamoto, N. Ishizu, Cutting Edge of High-Entropy Alloy Superconductors from the Perspective of Materials Research, Metals 10 (2020), 1078.

[2] L. Sun, R. J. Cava, High-entropy alloy superconductors: Status, opportunities, and challenges, Phys. Rev. Mater. 3 (2019), 090301.

[3] J.-W. Yeh, Alloy Design Strategies and Future Trends in High-Entropy Alloys, JOM 65 (2013), 1759-1771.

[4] E. P. George, D. Raabe, R. O. Ritchie, High-entropy alloys, Nat. Rev. Mater. 4 (2019), 515-534.

[5] Y. F. Ye, Q. Wang, J. Lu, C. T. Liu, Y. Yang, High-entropy alloy: challenges and prospects, Mater. Today 19 (2016), 349-362.

[6] D. B. Miracle, O. N. Senkov, A critical review of high entropy alloys and related concepts, Acta Mater. 122 (2017), 448-511.

[7] B. Yang, Y. Zhang, H. Pan, W. Si, Q. Zhang, Z. Shen, Y. Yu, S. Lan, F. Meng, Y. Liu, H. Huang, J. He, L. Gu, S. Zhang, L. Q. Chen, J. Zhu, C. W. Nan, Y. H. Lin, High-entropy enhanced capacitive energy storage, Nat. Mater. 21 (2022), 1074-1080.

[8] J. Ma, C. Huang, High entropy energy storage materials: Synthesis and application, J. Energy Storage 66 (2023), 107419.

[9] A. Sarkar, L. Velasco, D. Wang, Q. Wang, G. Talasila, L. Biasi, C. Kübel, T. Brezesinski, S. S. Bhattacharya, H. Hahn, B. Breitung, High entropy oxides for reversible energy storage, Nat. Commun 9 (2018), 3400.

[10] Y. J. Zhou, Y. Zhang, Y. L. Wang, G. L. Chen, Microstructure and compressive properties of multicomponent $Al_x(TiVCrMnFeCoNiCu)_{100-x}$ high-entropy alloys， Mater. Sci. Eng. A 454-455 (2007), 260-265.

[11] Y. Zou, H. Ma, R. Spolenak, Ultrastrong ductile and stable high-entropy alloys at small scales, Nat. Commun 6 (2015), 7748.

[12] L. Sun, Y. Luo, X. Ren, Z. Gao, T. Du, Z. Wu, J. Wang, A multicomponent $\gamma$-type $(Gd_{1/6}Tb_{1/6}Dy_{1/6}Tm_{1/6}Yb_{1/6}Lu_{1/6})_2Si_2O_7$ disilicate with outstanding thermal stability, Mater. Res. Lett. 8 (2020), 424-430.

[13] Y. Shi, B. Yang, P. K. Liaw, Metals 7 (2017), 43.

[14] L. Zeng, X. Hu, M. Boubeche, K. Li, L. Li, P. Yu, K. Wang, C. Zhang, K. Jin, D. Yao, H. Luo, Extremely strong coupling *s*-wave superconductivity in the medium-entropy alloy TiHfNbTa, Sci. China Phys. Mech. 66 (2023), 277412.

[15] P. Koželj, S. Vrtnik, A. Jelen, S. Jazbec, Z. Jagličić, S. Maiti, M. Feuerbacher, W. Steurer, J. Dolinšek, Discovery of a Superconducting High-Entropy Alloy, Phys. Rev. Lett. 113 (2014), 107001.

[16] K. Li, X. Hu, R. Guo, W. Jiang, L. Zeng, L. Li, P. Yu, K. Wang, C. Zhang, S. Guo, R. Zhong, T. Xie, D. Yao, H. Luo, Superconductivity in the Medium-Entropy Alloy TiVNbTa with a Body-Centered Cubic Structure, J. Phys. Chem. C 127 (2023), 16211-16218.

[17] S. A. Uporov, R. E. Ryltsev, V. A. Sidorov, S. K. Estemirova, E. V. Sterkhov, I. A. Balyakin, N. M. Chtchelkatchev, Pressure effects on electronic structure and electrical conductivity of TiZrHfNb high-entropy alloy, Intermetallics 140 (2022), 107394.

[18] S. Gabáni, J. Cedervall, G. Ek, G. Pristáš, M. Orendáč, J. Bačkai, O. Onufriienko, E. Gažo, K. Flachbart, Search for superconductivity in hydrides of TiZrNb, TiZrNbHf and TiZrNbHfTa equimolar alloys, Physica B 648 (2023), 414414.

[19] W. L. Nelson, A. T. Chemey, M. Hertz, E. Choi, D. E. Graf, S. Latturner, T. E. Albrecht-Schmitt, K. Wei, R. E. Baumbach, Baumbach, Superconductivity in a uranium containing high entropy alloy, Sci. Rep. 10 (2020), 4717.

[20] M. Parizh, Y. Lvovsky, M. Sumption, Conductors for commercial MRI magnets beyond NbTi: requirements and challenges, Supercond. Sci. Tech. 30 (2017), 014007.

[21] J.-F. Zhang, M. Gao, K. Liu, Z.-Y. Lu, First-principles study of the robust superconducting state of NbTi alloys under ultrahigh pressures, Phys. Rev. B 102 (2020), 195140.

[22] J. Guo, G. Lin, S. Cai, C. Xi, C. Zhang, W. Sun, Q. Wang, K. Yang, A. Li, Q. Wu, Y. Zhang, T. Xiang, R. J. Cava, L. Sun, Adv. Mater. 31 (2019), 1807240.

[23] H. Luo, W. Xie, J. Tao, H. Inoue, A. Gyenis, J. W. Krizan, A. Yazdani, Y. Zhu, R. J. Cava, Polytypism, polymorphism, and superconductivity in $TaSe_{2-x}Te_x$, Proc. Natl. Acad. Sci. 112 (2015), E1174-E1180.

[24] A. M. Clogston, Upper Limit for the Critical Field in Hard Superconductors Phys. Rev. Lett. 9 (1962), 266-267.

[25] W. L. McMillan, Transition Temperature of Strong-Coupled Superconductors, Phys. Rev. 167 (1968), 331-344.

[26] L. Zeng, X. Hu, S. Guo, G. Lin, J. Song, K. Li, Y. He, Y. Huang, C. Zhang, P. Yu, J. Ma, D. Yao, H. Luo, $Ta_4CoSi$: A tantalum-rich superconductor with a honeycomb network structure, Phys. Rev. B 106 (2022), 134501.

[27] L. Zeng, J. Zhan, M. Boubeche, K. Li, L. Li, P. Yu, K. Wang, C. Zhang, K. Jin, Y. Sun, H. Luo, Superconductivity in the bcc-Type High-Entropy Alloy TiHfNbTaMo, Adv Quantum Technol. n/a (2023) 2300213.

[28] N. Ishizu, J. Kitagawa, New high-entropy alloy superconductor $Hf_{21}Nb_{25}Ti_{15}V_{15}Zr_{24}$, Results Phys. 13 (2019), 102275.

[29] F. von Rohr, M. J. Winiarski, J. Tao, T. Klimczuk, R. J. Cava, Effect of electron count and chemical complexity in the Ta-Nb-Hf-Zr-Ti high-entropy alloy superconductor, Proc. Natl. Acad. Sci. 113 (2016), E7144-E7150.

[30] K. Jasiewicz, B. Wiendlocha, K. Górnicka, K. Gofryk, M. Gazda, T. Klimczuk, Pressure effects on the electronic structure and superconductivity of $(TaNb)_{0.67}(HfZrTi)_{0.33}$ high entropy alloy J. Tobola, Phys. Rev. B 100 (2019), 184503.

[31] K.-Y. Wu, S.-K. Chen, J.-M. Wu, Superconducting in equal molar NbTaTiZr-based high-entropy alloys, Nat. Sci. 10 (2018), 110.

[32] K. Motla, P. Meena, D. Singh, P. Biswas, A. Hillier, R. Singh, Superconducting and normal-state properties of the high-entropy alloy Nb-Re-Hf-Zr-Ti investigated by muon spin relaxation and rotation, Phys. Rev. B 105 (2022), 144501.

[33] J. H. Kim, R. Hidayati, S.-G. Jung, Y. A. Salawu, H.-J. Kim, J. H. Yun, J.-S. Rhyee, Enhancement of critical current density and strong vortex pinning in high entropy alloy superconductor $Ta_{1/6}Nb_{2/6}Hf_{1/6}Zr_{1/6}Ti_{1/6}$ synthesized by spark plasma sintering, Acta Mater. 232 (2022), 117971.

[34] B. T. Matthias, Empirical Relation between Superconductivity and the Number of Valence Electrons per Atom, Phys. Rev. 97 (1955), 74-76.

[35] M. M. Collver, R. H. Hammond, Superconductivity in "Amorphous" Transition-Metal Alloy Films, Phys. Rev. Lett. 30 (1973), 92-95.

[36] W. Nowak, M. Babij, A. Hanc-Kuczkowska, P. Sobota, A. Pikul, R. Idczak, Effect of the Presence of Structural Defects on the Superconducting Properties of $(NbTa)_{0.67}(MoHfW)_{0.33}$ and Nb-47wt%Ti, Metals 13 (2023), 1779.

[37] Z. An, S. Mao, Y. Liu, L. Wang, H. Zhou, B. Gan, Z. Zhang, X. Han, A novel HfNbTaTiV high-entropy alloy of superior mechanical properties designed on the principle of maximum lattice distortion, J. Mater. Sci. Technol. 79 (2021), 109-117.

**Superconductivity of the New Medium-Entropy Alloy $V_4Ti_2W$ with a Body-Centered Cubic Structure**

*Kuan Li*[a], *Weijie Lin*[b,c], *Ruixin Guo* [b,c], *Shu Guo*[b,c], *Lingyong Zeng*[a], *Longfu Li*[a], *Peifeng Yu*[a], *Kangwang Wang*[a], *Chao Zhang*[a], *Huixia Luo*[a*]

[a] School of Materials Science and Engineering, State Key Laboratory of Optoelectronic Materials and Technologies, Guangdong Provincial Key Laboratory of Magnetoelectric Physics and Devices, Key Lab of Polymer Composite & Functional Materials, Sun Yat-Sen University, No. 135, Xingang Xi Road, Guangzhou, 510275, P. R. China

[b] Shenzhen Institute for Quantum Science and Engineering, Southern University of Science and Technology, Shenzhen 518055, China;

[c] International Quantum Academy, Shenzhen 518048, China

**Table S1.** The $T_c$, VEC, and $\Delta S_{mix}$/R of the BCC-type MEAs and HEAs with $T_c$ higher than 5 K. [1-12]

| bcc-type MEAs and HEAs | $T_c$ | VEC | $\Delta S_{mix}$/R |
|---|---|---|---|
| $(TaNb)_{0.3}(HfZrTi)_{0.7}$ | 5.21 K | 4.30 | 1.47 |
| $(TaNb)_{0.4}(HfZrTi)_{0.6}$ | 6.25 K | 4.40 | 1.65 |
| $(TaNb)_{0.5}(HfZrTi)_{0.5}$ | 6.42 K | 4.50 | 1.73 |
| $(TaNb)_{0.55}(HfZrTi)_{0.45}$ | 6.68 K | 4.55 | 1.74 |
| $(TaNb)_{0.6}(HfZrTi)_{0.4}$ | 7.17 K | 4.60 | 1.71 |
| $(TaNb)_{0.65}(HfZrTi)_{0.35}$ | 7.49 K | 4.65 | 1.66 |
| $(TaNb)_{0.67}(HfZrTi)_{0.33}$ | 7.75 K | 4.67 | 1.64 |
| $(TaNb)_{0.7}(HfZrTi)_{0.3}$ | 8.03 K | 4.70 | 1.58 |
| $(TaNb)_{0.75}(HfZrTi)_{0.25}$ | 7.81 K | 4.75 | 1.47 |
| $(TaNb)_{0.8}(HfZrTi)_{0.2}$ | 7.57 K | 4.79 | 1.33 |
| $(TaV)_{0.67}(HfZrTi)_{0.33}$ | 7.20 K | 4.67 | 1.37 |
| $Hf_{21}Nb_{25}Ti_{15}V_{15}Zr_{24}$ | 5.30 K | 4.4 | 1.59 |
| $Nb_{21.5}Ta_{18.1}Ti_{15.9}Zr_{14.4}Hf_{16.6}V_{13.5}$ | 5.1 K | 4.53 | 1.78 |
| $Nb_{20}Re_{20}Zr_{20}Hf_{20}Ti_{20}$ | 5.3 K | 4.8 | 1.61 |
| $V_4Ti_2W$ (This work) | 5.57 K | 4.86 | 0.96 |
| TiHfNbTa | 6.75 K | 4.5 | 1.39 |
| $Nb_{22.1}Ta_{26.3}Ti_{16.6}Zr_{15.5}Hf_{19.5}$ | 7.1 K | 4.48 | 1.59 |
| $(NbV)_{0.67}(HfZrTi)_{0.33}$ | 7.2 K | 4.67 | 1.65 |
| $Ta_{1/6}Nb_{2/6}Hf_{1/6}Zr_{1/6}Ti_{1/6}$ | 7.9 K | 4.5 | 1.56 |
| NbTaTiZrGeV | 8.4 K | 4.5 | 1.79 |

## REFERENCES

[1] L. Sun, R. J. Cava, High-entropy alloy superconductors: Status, opportunities, and challenges, Phys. Rev. Mater. 3 (2019), 090301.

[2] P. Koželj, S. Vrtnik, A. Jelen, S. Jazbec, Z. Jagličić, S. Maiti, M. Feuerbacher, W. Steurer, J. Dolinšek, Discovery of a Superconducting High-Entropy Alloy, Phys. Rev. Lett. 113 (2014), 107001.

[3] K. Li, X. Hu, R. Guo, W. Jiang, L. Zeng, L. Li, P. Yu, K. Wang, C. Zhang, S. Guo, R. Zhong, T. Xie, D. Yao, H. Luo, Superconductivity in the Medium-Entropy Alloy TiVNbTa with a Body-Centered Cubic Structure, J. Phys. Chem. C 127 (2023), 16211-16218.

[4] L. Zeng, X. Hu, M. Boubeche, K. Li, L. Li, P. Yu, K. Wang, C. Zhang, K. Jin, D. Yao, H. Luo, Extremely strong coupling *s*-wave superconductivity in the medium-entropy alloy TiHfNbTa, Sci. China Phys. Mech. 66 (2023), 277412.

[5] N. Ishizu, J. Kitagawa, New high-entropy alloy superconductor $Hf_{21}Nb_{25}Ti_{15}V_{15}Zr_{24}$, Results Phys. 13 (2019), 102275.

[6] F. von Rohr, M. J. Winiarski, J. Tao, T. Klimczuk, R. J. Cava, Effect of electron count and chemical complexity in the Ta-Nb-Hf-Zr-Ti high-entropy alloy superconductor, Proc. Natl. Acad. Sci. 113 (2016), E7144-E7150.

[7] K. Jasiewicz, B. Wiendlocha, K. Górnicka, K. Gofryk, M. Gazda, T. Klimczuk, Pressure effects on the electronic structure and superconductivity of $(TaNb)_{0.67}(HfZrTi)_{0.33}$ high entropy alloy J. Tobola, Phys. Rev. B 100 (2019), 184503.

[8] K.-Y. Wu, S.-K. Chen, J.-M. Wu, Superconducting in equal molar NbTaTiZr-based high-entropy alloys, Nat. Sci. 10 (2018), 110.

[9] K. Motla, P. Meena, D. Singh, P. Biswas, A. Hillier, R. Singh, Superconducting and normal-state properties of the high-entropy alloy Nb-Re-Hf-Zr-Ti investigated by muon spin relaxation and rotation, Phys. Rev. B 105 (2022), 144501.

[10] J. H. Kim, R. Hidayati, S.-G. Jung, Y. A. Salawu, H.-J. Kim, J. H. Yun, J.-S. Rhyee, Enhancement of critical current density and strong vortex pinning in high entropy alloy superconductor $Ta_{1/6}Nb_{2/6}Hf_{1/6}Zr_{1/6}Ti_{1/6}$ synthesized by spark plasma sintering, Acta Mater. 232 (2022), 117971.

[11] W. Nowak, M. Babij, A. Hanc-Kuczkowska, P. Sobota, A. Pikul, R. Idczak, Effect of the Presence of Structural Defects on the Superconducting Properties of $(NbTa)_{0.67}(MoHfW)_{0.33}$ and Nb-47wt%Ti, Metals 13 (2023), 1779.

[12] Z. An, S. Mao, Y. Liu, L. Wang, H. Zhou, B. Gan, Z. Zhang, X. Han, A novel HfNbTaTiV high-entropy alloy of superior mechanical properties designed on the principle of maximum lattice distortion, J. Mater. Sci. Technol. 79 (2021), 109-117.